\documentclass[epjCONF]{svjour}
\usepackage{graphics}
\usepackage[varg]{txfonts} 
\usepackage[latin1]{inputenc}
\session-title{The Space Photometry Revolution}

 \def\aj{{AJ~}}  
\def\apj{{ApJ~}}                 
\def\apjl{{ApJ~}}                
   
\def\nat{{Nature~}}    
\def\mnras{{MNRAS~}}

\begin{document}
\title{What asteroseismology can do for exoplanets}
\author{Vincent Van Eylen\inst{1,2}\fnmsep\thanks{\email{vincent@phys.au.dk}} \and Mikkel N. Lund\inst{1,4} \and Victor Silva Aguirre\inst{1} \and Torben Arentoft\inst{1} \and Hans Kjeldsen\inst{1} \and Simon Albrecht\inst{1,3} \and William J. Chaplin\inst{5} \and Howard Isaacson\inst{6} \and May G. Pedersen\inst{1} \and Jens Jessen-Hansen\inst{1} \and Brandon Tingley\inst{1} \and J\o rgen Christensen-Dalsgaard\inst{1} \and Conny Aerts\inst{2} \and Tiago L. Campante\inst{5} \and Steve T. Bryson\inst{7}
}


%
\institute{Stellar Astrophysics Centre, Department of Physics and Astronomy, Aarhus University, Ny Munkegade 120,
DK-8000 Aarhus C, Denmark \and
Instituut voor Sterrenkunde, Katholieke Universiteit Leuven, Celestijnenlaan 200 B, B-3001 Heverlee, Belgium\label{inst2} \and
Department of Physics, and Kavli Institute for Astrophysics and Space Research, Massachusetts Institute of Technology, Cambridge, MA 02139, USA\label{inst3} \and
Sydney Institute for Astronomy (SIfA), School of Physics, University of Sydney, NSW 2006, Australia\label{inst4} \and 
School of Physics and Astronomy, University of Birmingham, Edgbaston, Birmingham, B15 2TT, UK\label{inst5} \and 
Department of Astronomy, University of California, Berkeley, CA 94820, USA\label{inst6} \and 
NASA Ames Research Center, Moffett Field, CA 94035, USA }

%
\abstract{
We describe three useful applications of asteroseismology in the context of exoplanet science: (1) the detailed characterisation of exoplanet host stars; (2) the measurement of stellar inclinations; and (3) the determination of orbital eccentricity from transit duration making use of asteroseismic stellar densities. We do so using the example system Kepler-410 \cite{vaneylen2014}. This is one of the brightest (V = 9.4) \textit{Kepler} exoplanet host stars, containing a small (2.8 $R_\oplus$) transiting planet in a long orbit (17.8 days), and one or more additional non-transiting planets as indicated by transit timing variations. The validation of Kepler-410 (KOI-42) was complicated due to the presence of a companion star, and the planetary nature of the system was confirmed after analyzing a \textit{Spitzer} transit observation as well as ground-based follow-up observations.
%
} 

\maketitle
\section{Introduction}
\label{sec:introduction}

Asteroseismology can be of exceptional use in understanding and characterising planetary systems. It is well known in exoplanet science that a deep understanding of an extrasolar planet requires knowledge of the host star. We argue that in the context of exoplanets, asteroseismology can make three major contributions:

\begin{enumerate}
 \item precise characterisation (i.e. mass, radius, age, ...) of exoplanet host stars;
 \item obliquity determination of exoplanet systems by measuring stellar inclinations;
 \item exoplanet eccentricity determination (without radial velocity measurements), by comparing stellar densities with measurements of transit durations.
\end{enumerate}

\begin{figure}
\centering
\resizebox{0.6\columnwidth}{!}{%
\includegraphics{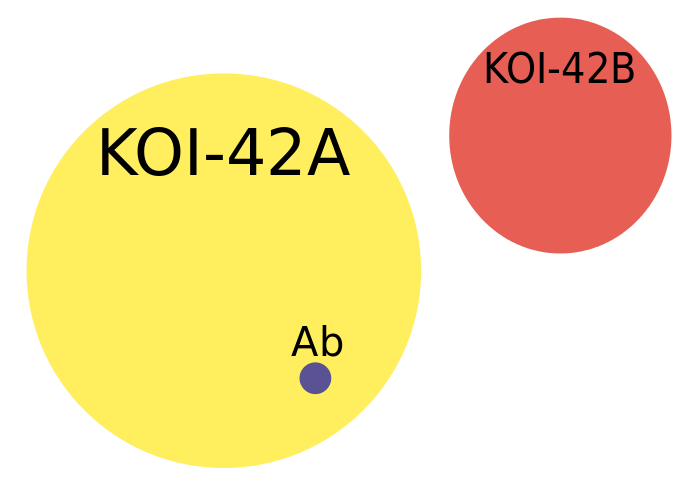}
\includegraphics{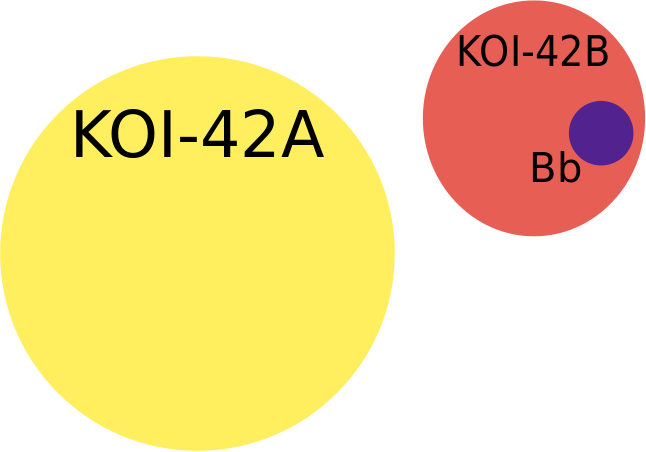}}
\resizebox{0.6\columnwidth}{!}{
 \includegraphics{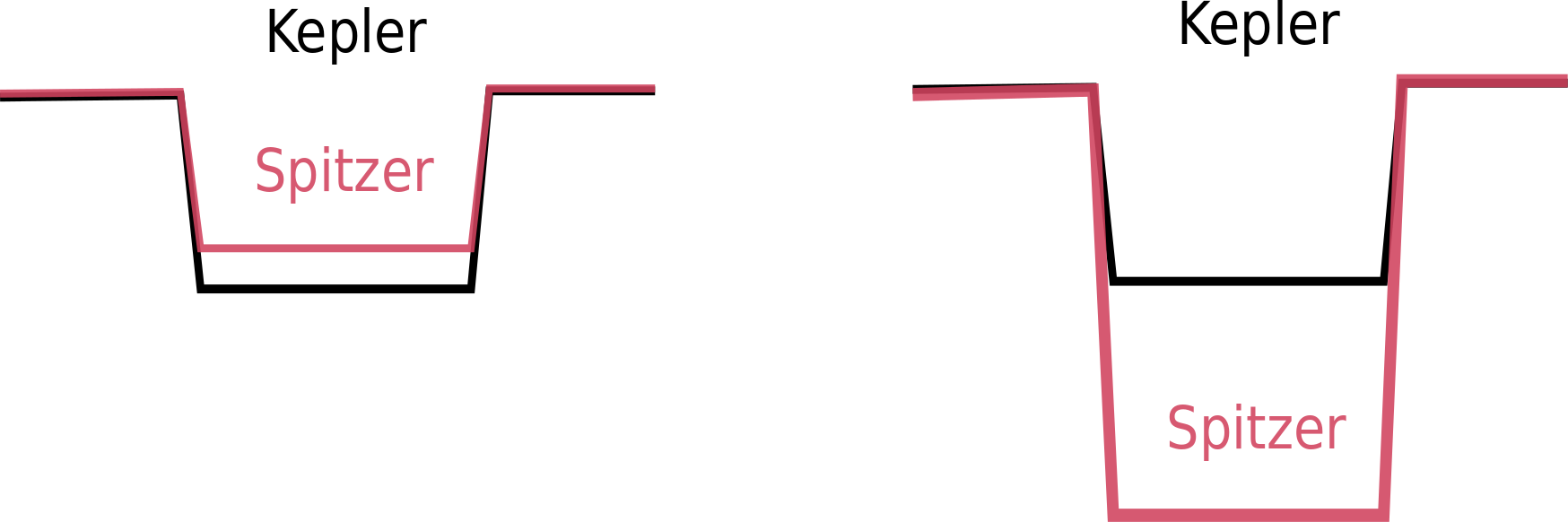} }
\caption{Two scenarios, a planet around Kepler-410A or a planet around KOI-410B, provide identical transit depths in the \textit{Kepler} band. However, a different transit depth is anticipated in Spitzer observations because of the different colors of KOI-42A and KOI-42B. Spitzer observations (see \cite{vaneylen2014}) decisively show that the planet orbits around Kepler-410A (left figure).}
\label{fig:transitdepthboth}       
\end{figure}

To illustrate these points, we make use of the example system Kepler-410 \cite{vaneylen2014} (formerly KOI-42), one of the brightest (V = 9.4) \textit{Kepler} planet hosts. The system consists of two stars, Kepler-410A and Kepler-410B, which are indistinguishable in \textit{Kepler} photometry but have been resolved in ground-based follow-up observations \cite{howell2011,adams2012}. The planet candidate (KOI-42b) was firmly established to be a planet orbiting Kepler-410A, making use of multi-color photometry, i.e.\ the use of \textit{Spitzer} $4.5~\mu$m observations. The data were analyzed in \cite{vaneylen2014} and show a \textit{Spitzer} transit depth which is essentially equal to the \textit{Kepler} depth, as would be expected for a transiting planet orbiting Kepler-410A, while a planet orbiting Kepler-410B would cause a far deeper \textit{Spitzer} transit because the different stellar energy distributions change the relative contributions of the two stars with different wavelengths. This principle is 
illustrated in Figure~\ref{fig:transitdepthboth}. For more details on the planetary validation, we refer to \cite{vaneylen2014}.\\

In the following we refer to the planet as Kepler-410A b. Its orbital period is 17.833648 $\pm$ 0.000054 days and its radius is 2.838 $\pm$ 0.054 $R_\oplus$ \cite{vaneylen2014}. Transit timing variations indicate Kepler-410 to be a multi-planet system, as they reveal the presence of at least one additional planet orbiting Kepler-410A.

%

\section{What asteroseismology can do}

\subsection{Precise host star characterisation}

In the context of exoplanets asteroseismology is perhaps most known for its ability to characterize stellar properties to a high level of precision. Stellar pulsations influence the brightness of stars, and the quality of \textit{Kepler} time series allows for the extraction of oscillation frequencies after creating power spectra using Fourier transformation. Stellar oscillations' global patterns, such as the large frequency separation and frequency of maximum power, have been used to extract masses and radii of samples of exoplanet host stars (e.g.\ \cite{huber2013catalog}). Detailed modeling of individual frequencies has been done for several interesting exoplanet hosts, such as Kepler-10 \cite{batalha2011,fogtmannschulz2014} and HAT-P-7 \cite{christensendalsgaard2010,vaneylen2012,lund2014}. For Kepler-410, the stellar radius was measured to the percentage level \cite{vaneylen2014}, allowing a similar precision to be achieved for the planetary radius, a level at which systematics in the \textit{Kepler} 
observations might become apparent \cite{vaneylen2013}. 

\subsection{Stellar inclination and obliquity}
In cases with exceptionally high signal to noise photometric measurements, effects in the stellar oscillations caused by stellar rotation can be measured. Such \textit{rotational splittings} provide information on the stellar rotation rate, but the relative visibilities of the modes in triplets and quintuplets also reveal the stellar inclination \cite{gizon2003}. This is because these visibilities are affected by geometric cancellation effects which depend on the stellar inclination. The latter is of particular use for exoplanet host stars, as it can reveal the obliquity of such systems.\\

\begin{figure}
\resizebox{0.9\columnwidth}{!}{%
\includegraphics{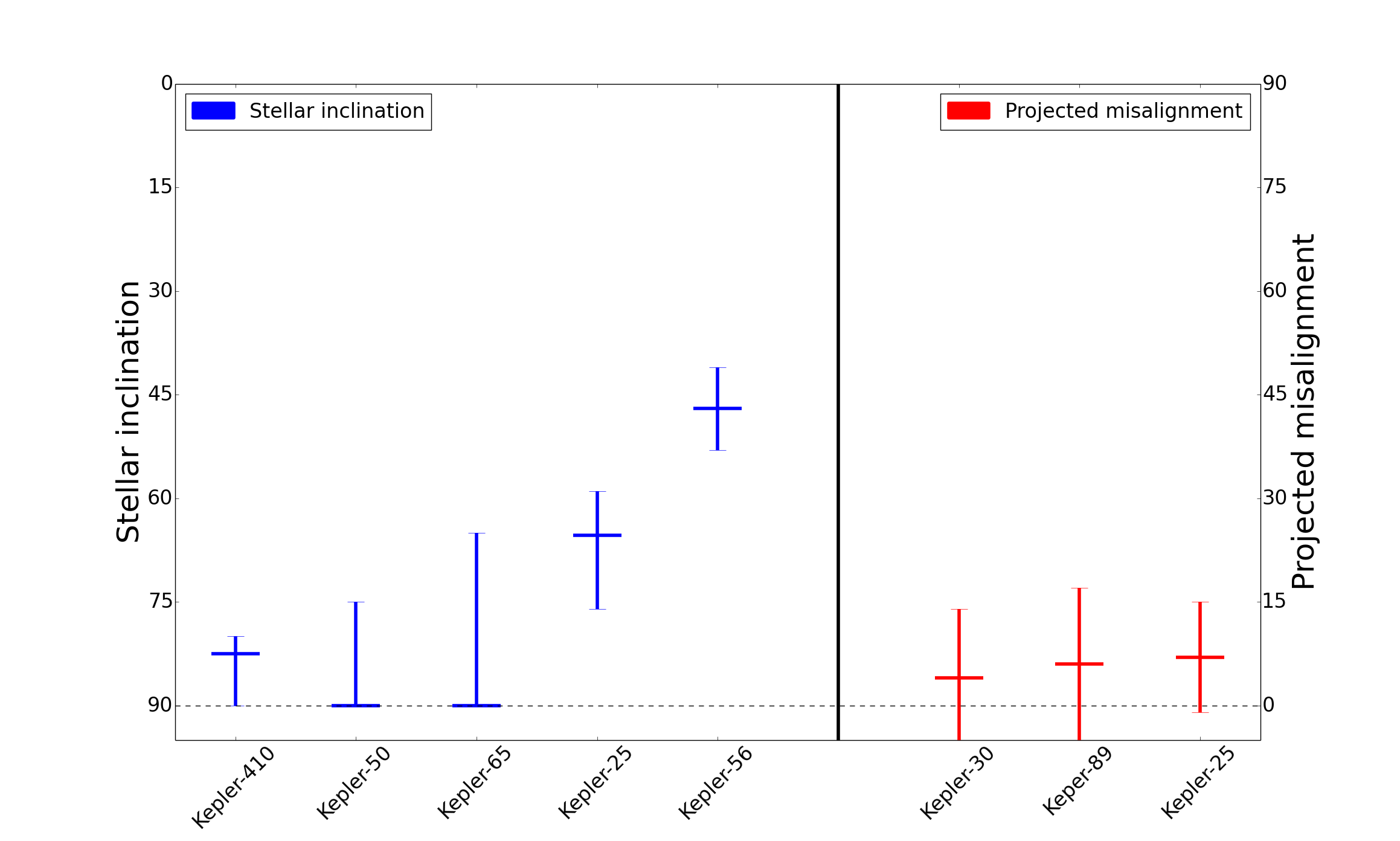}[!bh]
}
\caption{Stellar inclinations and (mis)alignments of multi-planet systems. Stellar inclinations were measured using asteroseismology and taken from \cite{vaneylen2014} (Kepler-410), \cite{chaplin2013} (Kepler-50 and Kepler-65), \cite{benomar2014} (Kepler-25) and \cite{huber2013} (Kepler-56). Projected misalignments are measured using spot crossing events \cite{sanchisojeda2012} (Kepler-30) and Rossiter-McLaughlin measurements \cite{hirano2012,albrecht2013} (Kepler-89 and Kepler-25).}
\label{fig:obliquity_overview}       
\end{figure}

The obliquities of Hot Jupiters display a wide variety \cite{winn2010,albrecht2012}, as determined from Rossiter-Mc\-Laugh\-lin observations. One possible explanation for the observed misalignments is that they have occurred during planetary migration (e.g.\ \cite{rasio1996,fabrycky2007}), while another explanation assumes primordial star-disk misalignment (e.g.\ \cite{bate2010,batygin2012}). Observations of obliquities in multi-planet systems provide the possibility to distinguish between these scenarios. Measurements of obliquities in multi-planet systems, which typically have smaller planets, are however challenging using Rossiter-McLaughlin measurements, which are dependent on the transit depth. Asteroseismology, which directly measures stellar inclinations, is independent of the planetary size. A handful of multi-planet systems have currently been measured (see Figure~\ref{fig:obliquity_overview}). When compared to similar figures for Hot Jupiters (e.g.\ \cite{albrecht2012}), the obliquity distribution 
of multi-planet systems seem flatter (more aligned), consistent with \cite{albrecht2013} and \cite{morton2014}. However, the sample remains small.

\section{Orbital eccentricity}

Transit durations, \textit{assuming circular orbits}, provide a measurement for the relative semi-major axis, $a/R_\star$, which (for a known orbital period) can be converted into the stellar density using Kepler's third law \cite{seager2003}. If an independent measurement of the stellar density is available (e.g.\ from asteroseismology), then the transit duration can be used to test for circularity of the planetary orbit \cite{tingley2011}, as a non-zero eccentricity would change the transit duration. In case of disagreement in stellar density estimates, the anomalous transit duration can be related to the orbital eccentricity, which requires taking into account an (unknown) angle of periastron. Several other effects can complicate the measurement, e.g.\ transit timing variations, light curve blending, ... \cite{kipping2014}.\\

The method has been used to measure the eccentricity of Jupiter-sized planets \cite{dawson2012,dawson2014}. For Kepler-410A b, we find a low but significantly non-zero eccentricity (0.17$^{+0.07}_{-0.06}$) \cite{vaneylen2014}. We recall that Kepler-410A b, a super-Earth in a 17.8 day orbit, is beyond the range of eccentricity measurements using radial velocities, illustrating the potential of this technique in measuring eccentricities for samples of small planets \cite{vaneylen2015}.\\


\begin{thebibliography}{29}

\bibitem{vaneylen2014}
V.~{Van Eylen}, M.N. {Lund}, V.~{Silva Aguirre}, T.~{Arentoft}, H.~{Kjeldsen},
  S.~{Albrecht}, W.J. {Chaplin}, H.~{Isaacson}, M.G. {Pedersen},
  J.~{Jessen-Hansen} et~al., \apj \textbf{782}, 14 (2014)

\bibitem{howell2011}
S.B. {Howell}, M.E. {Everett}, W.~{Sherry}, E.~{Horch}, D.R. {Ciardi}, \aj
  \textbf{142}, 19 (2011)

\bibitem{adams2012}
E.R. {Adams}, D.R. {Ciardi}, A.K. {Dupree}, T.N. {Gautier}, III, C.~{Kulesa},
  D.~{McCarthy}, \aj \textbf{144}, 42 (2012)

\bibitem{huber2013catalog}
D.~{Huber}, W.J. {Chaplin}, J.~{Christensen-Dalsgaard}, R.L. {Gilliland},
  H.~{Kjeldsen}, L.A. {Buchhave}, D.A. {Fischer}, J.J. {Lissauer}, J.F. {Rowe},
  R.~{Sanchis-Ojeda} et~al., \apj \textbf{767}, 127 (2013)

\bibitem{batalha2011}
N.M. {Batalha}, W.J. {Borucki}, S.T. {Bryson}, L.A. {Buchhave}, D.A.
  {Caldwell}, J.~{Christensen-Dalsgaard}, D.~{Ciardi}, E.W. {Dunham},
  F.~{Fressin}, T.N. {Gautier}, III et~al., \apj \textbf{729}, 27 (2011)

\bibitem{fogtmannschulz2014}
A.~{Fogtmann-Schulz}, B.~{Hinrup}, V.~{Van Eylen}, J.~{Christensen-Dalsgaard},
  H.~{Kjeldsen}, V.~{Silva Aguirre}, B.~{Tingley}, \apj \textbf{781}, 67
  (2014)

\bibitem{christensendalsgaard2010}
J.~{Christensen-Dalsgaard}, H.~{Kjeldsen}, T.M. {Brown}, R.L. {Gilliland},
  T.~{Arentoft}, S.~{Frandsen}, P.O. {Quirion}, W.J. {Borucki}, D.~{Koch}, J.M.
  {Jenkins}, \apjl \textbf{713}, L164 (2010)

\bibitem{vaneylen2012}
V.~{Van Eylen}, H.~{Kjeldsen}, J.~{Christensen-Dalsgaard}, C.~{Aerts},
  Astronomische Nachrichten \textbf{333}, 1088 (2012)

\bibitem{lund2014}
M.N. {Lund}, M.~{Lundkvist}, V.~{Silva Aguirre}, G.~{Houdek}, L.~{Casagrande},
  V.~{Van Eylen}, T.L. {Campante}, C.~{Karoff}, H.~{Kjeldsen}, S.~{Albrecht}
  et~al., ArXiv e-prints  (2014)

\bibitem{vaneylen2013}
V.~{Van Eylen}, M.~{Lindholm Nielsen}, B.~{Hinrup}, B.~{Tingley},
  H.~{Kjeldsen}, \apjl \textbf{774}, L19 (2013)

\bibitem{gizon2003}
L.~{Gizon}, S.K. {Solanki}, \apj \textbf{589}, 1009 (2003)

\bibitem{chaplin2013}
W.J. {Chaplin}, R.~{Sanchis-Ojeda}, T.L. {Campante}, R.~{Handberg},
  D.~{Stello}, J.N. {Winn}, S.~{Basu}, J.~{Christensen-Dalsgaard}, G.R.
  {Davies}, T.S. {Metcalfe} et~al., \apj \textbf{766}, 101 (2013)

\bibitem{benomar2014}
O.~{Benomar}, K.~{Masuda}, H.~{Shibahashi}, Y.~{Suto}, ArXiv e-prints  (2014)

\bibitem{huber2013}
D.~{Huber}, J.A. {Carter}, M.~{Barbieri}, A.~{Miglio}, K.M. {Deck}, D.C.
  {Fabrycky}, B.T. {Montet}, L.A. {Buchhave}, W.J. {Chaplin}, S.~{Hekker}
  et~al., Science \textbf{342}, 331 (2013)

\bibitem{sanchisojeda2012}
R.~{Sanchis-Ojeda}, D.C. {Fabrycky}, J.N. {Winn}, T.~{Barclay}, B.D. {Clarke},
  E.B. {Ford}, J.J. {Fortney}, J.C. {Geary}, M.J. {Holman}, A.W. {Howard}
  et~al., \nat \textbf{487}, 449 (2012)

\bibitem{hirano2012}
T.~{Hirano}, R.~{Sanchis-Ojeda}, Y.~{Takeda}, N.~{Narita}, J.N. {Winn},
  A.~{Taruya}, Y.~{Suto}, \apj \textbf{756}, 66 (2012)

\bibitem{albrecht2013}
S.~{Albrecht}, J.N. {Winn}, G.W. {Marcy}, A.W. {Howard}, H.~{Isaacson}, J.A.
  {Johnson}, \apj \textbf{771}, 11 (2013)

\bibitem{winn2010}
J.N. {Winn}, D.~{Fabrycky}, S.~{Albrecht}, J.A. {Johnson}, \apjl \textbf{718},
  L145 (2010)

\bibitem{albrecht2012}
S.~{Albrecht}, J.N. {Winn}, J.A. {Johnson}, A.W. {Howard}, G.W. {Marcy}, R.P.
  {Butler}, P.~{Arriagada}, J.D. {Crane}, S.A. {Shectman}, I.B. {Thompson}
  et~al., \apj \textbf{757}, 18 (2012)

\bibitem{rasio1996}
F.A. {Rasio}, E.B. {Ford}, Science \textbf{274}, 954 (1996)

\bibitem{fabrycky2007}
D.~{Fabrycky}, S.~{Tremaine}, \apj \textbf{669}, 1298 (2007)

\bibitem{bate2010}
M.R. {Bate}, G.~{Lodato}, J.E. {Pringle}, \mnras \textbf{401}, 1505 (2010)

\bibitem{batygin2012}
K.~{Batygin}, \nat \textbf{491}, 418 (2012)

\bibitem{morton2014}
T.D. {Morton}, J.N. {Winn}, ArXiv e-prints  (2014)

\bibitem{seager2003}
S.~{Seager}, G.~{Mall{\'e}n-Ornelas}, \apj \textbf{585}, 1038 (2003)

\bibitem{tingley2011}
B.~{Tingley}, A.S. {Bonomo}, H.J. {Deeg}, \apj \textbf{726}, 112 (2011)

\bibitem{kipping2014}
D.M. {Kipping}, \mnras \textbf{440}, 2164 (2014)

\bibitem{dawson2012}
R.I. {Dawson}, J.A. {Johnson}, \apj \textbf{756}, 122 (2012)

\bibitem{dawson2014}
R.I. {Dawson}, J.A. {Johnson}, D.C. {Fabrycky}, D.~{Foreman-Mackey}, R.A.
  {Murray-Clay}, L.A. {Buchhave}, P.A. {Cargile}, K.I. {Clubb}, B.J. {Fulton},
  L.~{Hebb} et~al., \apj \textbf{791}, 89 (2014)

\bibitem{vaneylen2015}
V. {Van Eylen}, S. {Albrecht}, ArXiv:1505.02814
  
\end{thebibliography}

\end{document}